\documentclass[letterpaper]{article} 
\usepackage{aaai25}  
\usepackage{times}  
\usepackage{helvet}  
\usepackage{courier}  
\usepackage[hyphens]{url}  
\usepackage{graphicx} 
\usepackage{natbib}  
\usepackage{caption} 
\usepackage{algorithm}
\usepackage{algorithmic}
\usepackage{booktabs}
\usepackage{amsmath}
\usepackage{amssymb}
\usepackage{xcolor}
\usepackage{array}
\usepackage{cleveref}

\usepackage{newfloat}
\usepackage{listings}
\DeclareCaptionStyle{ruled}{labelfont=normalfont,labelsep=colon,strut=off} 
\floatstyle{ruled}
\newfloat{listing}{tb}{lst}{}
\floatname{listing}{Listing}
\nocopyright 

\title{Predicting Liquidity-Aware Bond Yields using Causal GANs and Deep Reinforcement Learning with LLM Evaluation}

\author{
    Jaskaran Singh Walia\textsuperscript{\rm 1}\equalcontrib,
    Aarush Sinha\textsuperscript{\rm 1}\equalcontrib,
    Naman Saraswat\textsuperscript{\rm 2}\equalcontrib,
    Srinitish Srinivasan\textsuperscript{\rm 1},
    Srihari Unnikrishnan\textsuperscript{\rm 1}
}
\affiliations{
    \textsuperscript{\rm 1}School of Computer Science and Engineering, Vellore Institute of Technology, India\\
    \textsuperscript{\rm 2}Indian Institute of Science, Bangalore, India
}

\usepackage{bibentry}

\begin{document}

\maketitle


\begin{abstract}
Financial bond yield forecasting is challenging due to data scarcity, nonlinear macroeconomic dependencies, and evolving market conditions. In this paper, we propose a novel framework that leverages Causal Generative Adversarial Networks (CausalGANs) and Soft Actor-Critic (SAC) reinforcement learning (RL) to generate high-fidelity synthetic bond yield data for four major bond categories (AAA, BAA, US10Y, Junk). By incorporating 12 key macroeconomic variables, we ensure statistical fidelity by preserving essential market properties. To transform this market dependent-synthetic data into actionable insights, we employ a fine-tuned Large Language Model (LLM) Qwen2.5-7B that generates trading signals (BUY/HOLD/SELL), risk assessments, and volatility projections. We use automated, human and LLM evaluations, all of which demonstrate that our framework improves forecasting performance over existing methods, with statistical validation via predictive accuracy, MAE evaluation(0.103\%), profit/loss evaluation (60\% profit rate), LLM evaluation (3.37/5) and expert assessments scoring 4.67 out of 5. The reinforcement learning-enhanced synthetic data generation achieves the least Mean Absolute Error of 0.103, demonstrating its effectiveness in replicating real-world bond market dynamics. We not only enhance data-driven trading strategies but also provides a scalable, high-fidelity synthetic financial data pipeline for risk \& volatility management and investment decision-making. This work establishes a bridge between synthetic data generation, LLM driven financial forecasting, and language model evaluation, contributing to AI-driven financial decision-making.

\end{abstract}


%

\section*{Introduction}

\begin{figure*}
    \centering
    \includegraphics[width=1\linewidth]{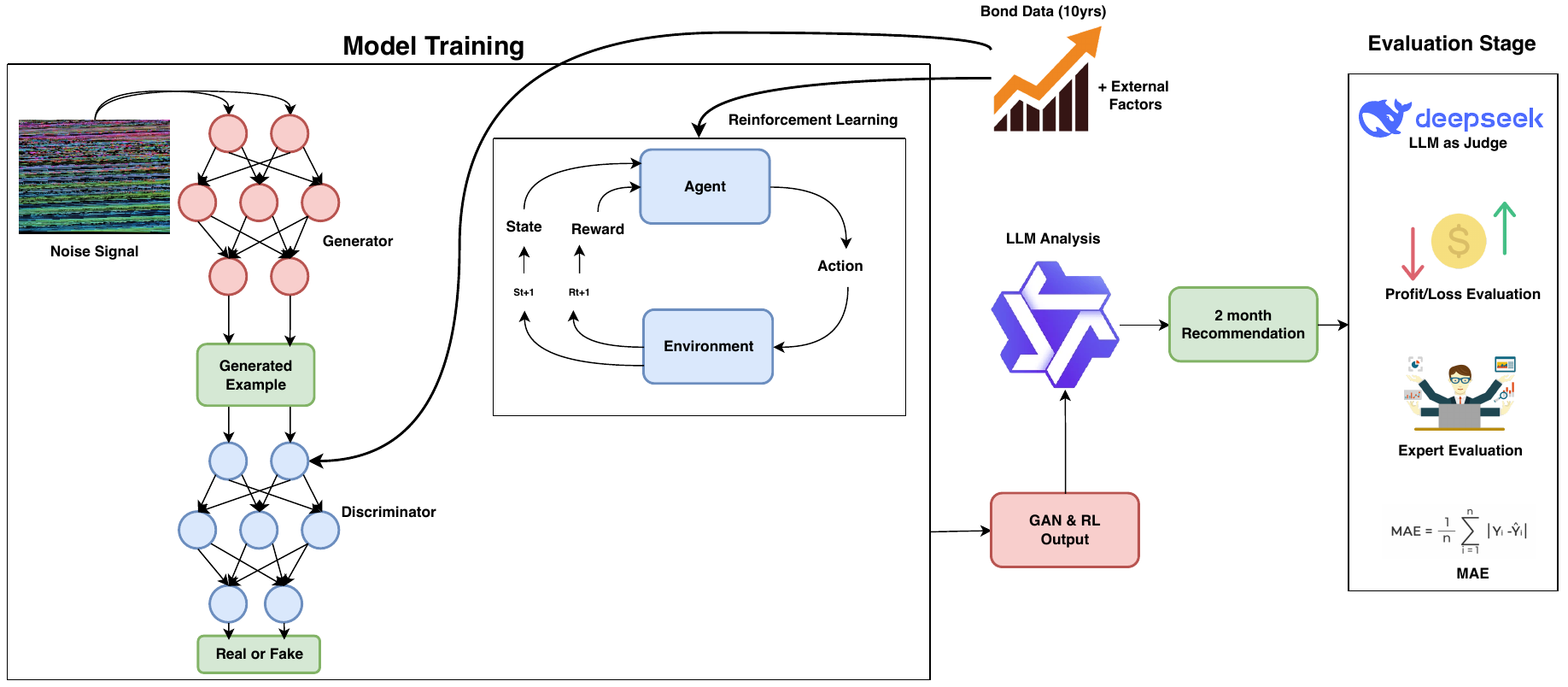}
    \caption{Overall pipeline of the training and evaluation stages of our proposed methodology.}
    \label{fig:pipeline}
\end{figure*}

Forecasting bond yields is a foundational challenge in financial markets, with significant implications for investment strategies, risk management, and monetary policy. The task is complicated by the nonlinear relationships between bond yields and macroeconomic variables, as well as data scarcity in less liquid categories such as high-yield junk bonds. Traditional models, including Nelson-Siegel\cite{yu2011forecasting} and autoregressive\cite{carriero2012forecasting} frameworks, often fail to capture these complexities, particularly during periods of market anomalies like inverted yield curves observed in the post-COVID era\cite{ghosh2021feb}. To address these challenges, we propose a novel framework that integrates Causal Generative Adversarial Networks (Causal GANs) with Soft Actor-Critic (SAC) reinforcement learning to generate high-fidelity synthetic bond yield data. By incorporating 12 key macroeconomic indicators, our approach ensures statistical fidelity while addressing data limitations and improving predictive accuracy.\cite{haarnoja2018soft}\cite{tamuly2024portfolio}

The practical utility of accurate bond yield forecasting extends beyond theoretical interest. Investors and portfolio managers rely on yield predictions to implement strategies such as duration management, yield curve positioning, and credit risk analysis\cite{bieri2005riding}. For example, strategies like the barbell approach—pairing short-term and long-term bonds—require a nuanced understanding of yield dynamics across maturities to optimize returns while mitigating risks\cite{trainor2020using}. Moreover, advanced techniques like roll-down strategies or sector rotations depend on precise forecasts to exploit changes in yields over time. Recognizing these needs, our research not only enhances predictive accuracy but also provides actionable trading signals (BUY/HOLD/SELL), risk assessments, and volatility projections through a fine-tuned Large Language Model (LLM), Qwen2.5-7B\cite{yang2024qwen2}

This research contributes to financial AI literature by developing: (1) the first application of RL \& Causal GANs for multi-class bond data synthesis, (2) an LLM architecture specifically optimized for fixed-income strategy formulation with integrated risk and volatility analytics, and (3) a unified evaluation protocol bridging statistical validation and economic rationality checks. Figure~\ref{fig:pipeline} illustrates the complete pipeline of our approach, which visually encapsulates the sequential process from data ingestion to final prediction. In the diagram, historical bond data for the 4 bonds are passed into our synthetic data generation module that employs a Causal GAN and parallelly to the RL module. The generated synthetic data (from both modules), alongside real data, are subsequently fed into our fine-tuned LLM for predictive analytics. Our experimental results demonstrate significant results in terms of profitability (where we demonstrate over a 60\% profit rate) and forecasting evaluated by an LLM (with a maximum score of 3.37/5 for RL) and human evaluation (with a max score of 4.67 for RL) both of which are higher than the prediction scores of the actual, non-generated data, which proves the efficacy of our method.

\section*{Related Work}

\subsection{GANs in Financial Data Generation}
Generative Adversarial Networks (GANs) have emerged as a versatile tool to overcome data scarcity and to create realistic synthetic datasets for financial applications. Early work by \cite{efimov2020using} demonstrated the feasibility of synthesizing artificial financial datasets from American Express data, paving the way for using GANs in benchmarking and research. Building on this, \cite{wiese2019quant} introduced Quant GANs, which leverage temporal convolutional networks to capture long-range dependencies—such as volatility clusters and leverage effects—in financial time series.
Other researchers have focused on modeling market microstructure. For instance, \cite{li2020generating} proposed a Stock-GAN that generates realistic stock market order streams by incorporating a conditional Wasserstein GAN and components mimicking the auction mechanism. In a related effort, \cite{rizzato2022generative} extended GAN-based methods to generate synthetic financial scenarios, addressing multivariate properties including price, market capitalization, and even ESG scores. Recent advances have also added control and context to the generative process. \cite{xia2023market} introduced Market-GAN, which integrates semantic context—such as market dynamics and stock tickers—to condition the generation of financial data. In parallel, attention mechanisms have been incorporated into GANs for time series simulation; \cite{fu2022simulating} employed attention to improve the reproduction of stylized facts like autocorrelation, while \cite{wiese2019deep} applied GANs to simulate equity option markets for deep hedging applications.

\subsection{Reinforcement Learning in Finance}
Recent years have witnessed a surge of interest in applying reinforcement learning (RL) techniques to financial problems. Early work such as \cite{zhang2019deep} introduced deep RL frameworks for trading by directly mapping market states to trading actions, demonstrating that end‐to‐end learned policies can outperform classical time series momentum strategies. In parallel, \cite{selser2021optimal} focused on market making, where an RL agent is trained to optimally set bid–ask quotes, balancing the tradeoff between capturing the spread and managing inventory risk. Another strand of research frames trading itself as a game. For example, \cite{huang2018financial} models financial trading as an interactive game and employs a deep recurrent Q-network to capture the dynamic and stochastic nature of market environments. Extending these ideas to portfolio optimization, \cite{yu2019model} proposes a model‐based RL approach that integrates prediction and control for dynamic asset allocation under transaction costs and risk considerations. Building on these foundations, \cite{li2021finrl} introduces the FinRL-Podracer framework, which emphasizes both high performance and scalability, thereby facilitating continuous training and integration into practical trading systems. Complementing these works, \cite{theate2020application} details an application of deep RL for algorithmic trading that incorporates novel reward functions to directly target risk-adjusted returns. Finally, survey work such as \cite{pippas2024evolution} synthesizes the rapidly growing literature by critically evaluating 167 publications on RL applications in finance. Recent contributions (\cite{huang2022deepreinforcementlearningportfolio, hambly2023recent}) further extend these ideas by addressing emerging challenges such as sample efficiency and risk-sensitive reward design in multi-agent settings. 

\subsection{LLMs in Financial Prediction and Evaluation}
Recent studies have demonstrated that large language models (LLMs) are rapidly transforming the landscape of financial prediction and evaluation. For instance, \cite{xie2023pixiulargelanguagemodel} introduces PIXIU, a comprehensive framework that not only fine-tunes a base LLM for financial tasks using a large-scale instruction dataset but also establishes an evaluation benchmark covering multiple financial domains. This work underlines the importance of domain-specific instruction data and standardized benchmarks in pushing forward the state of open-source financial AI. Building on this foundation, \cite{ding2023integratingstockfeaturesglobal} proposes an integration framework that combines traditional stock features with semantic representations derived from LLMs. Their Local-Global (LG) model, augmented by Self-Correlated Reinforcement Learning (SCRL), effectively aligns latent textual information with quantitative features, thereby enhancing stock return predictions. In a similar vein, \cite{wang2024llmfactorextractingprofitablefactors} presents LLMFactor, which utilizes Sequential Knowledge-Guided Prompting (SKGP) to extract interpretable factors that drive stock movement predictions. This approach not only improves explain ability but also provides clear insights into the dynamics of market changes. The predictive capabilities of LLMs have also been investigated from the perspective of raw model output. \cite{lopezlira2024chatgptforecaststockprice} explores whether ChatGPT can forecast stock price movements based solely on news headlines, finding that its sentiment scores significantly predict out-of-sample returns—even outperforming traditional quantitative models in certain settings. Complementing this, \cite{Fatouros_2024} develops MarketSenseAI, a framework that leverages GPT-4’s advanced reasoning (through chain-of-thought and in-context learning) for stock selection, demonstrating notable excess alpha and cumulative returns in competitive market universes. Recent efforts have also focused on the fine-tuning and adaptation of LLMs to financial time-series data. For example, \cite{guo2024finetuninglargelanguagemodels} compares encoder-only and decoder-only architectures in a fine-tuning setting using financial newsflow data, showing that token-level representations extracted from LLMs can serve as strong predictive signals for both long-only and long-short portfolio strategies. Moreover, \cite{kim2024financialstatementanalysislarge} investigates the use of LLMs for financial statement analysis, revealing that models such as GPT-4 not only outperform human analysts in directional earnings prediction but also yield trading strategies with superior risk-adjusted returns. Finally, \cite{Kirtac_2024} highlights the benefits of advanced sentiment analysis using LLMs. 

\subsection{Evaluation Frameworks in Financial AI}

The recent surge in applying large language models (LLMs) and other generative AI techniques to finance has prompted the development of specialized evaluation frameworks that capture the unique challenges and nuances of financial tasks. These frameworks assess general performance while also measuring domain-specific capabilities such as risk management, compliance, financial reasoning, and market behavior prediction. Several evaluation benchmarks have emerged, each targeting different facets of financial applications. For instance, SuperCLUE-Fin \cite{xu2024supercluefingradedfinegrainedanalysis} offers a graded, fine-grained analysis of Chinese financial LLMs by simulating real-life multi-turn financial conversations, evaluating models on logical reasoning, computational efficiency, and regulatory compliance. In another vein, frameworks like the one presented for generalist credit scoring \cite{feng2024empoweringmanybiasingfew} highlight the potential of LLMs as holistic evaluators in credit assessment, measuring not only accuracy but also biases and fairness in model predictions. Additional evaluation strategies focus on the relational and dynamic nature of financial data. For example, the framework for evaluating financial relational graphs \cite{niu2024evaluatingfinancialrelationalgraphs} assesses models' ability to explain historical financial phenomena by constructing and interpreting dynamic stock relationship graphs before integrating them into downstream predictive models. Complementing these approaches is FinBen \cite{xie2024finbenaholisticfinancialbenchmark}, a benchmark covering over two dozen financial tasks such as information extraction, textual analysis, forecasting, and decision-making, which evaluates state-of-the-art models tailored for stock trading and risk management. Finally, the emergence of frameworks like LOB-Bench \cite{nagy2025lobbenchbenchmarkinggenerativeai} underscores the importance of domain-specific metrics, particularly in financial markets, where the statistical nuances of order flows, price impacts, and market microstructure play a critical role. Collectively, these evaluation frameworks illustrate a trend toward increasingly sophisticated and nuanced assessments in financial AI, emphasizing the need for metrics that go beyond traditional accuracy to incorporate aspects such as fairness, interpretability, and robustness in high-stakes financial environments.

\begin{table*}[h]
    \centering
    \renewcommand{\arraystretch}{1.2}
    \small
    \begin{tabular}{>{\raggedright\arraybackslash}p{3cm}p{10cm}}
        \toprule
        \textbf{Variable} & \textbf{Effect on Bond Yields} \\
        \midrule
        Inflation Rate ($I_t$) & Higher inflation generally leads to higher bond yields as investors demand greater compensation for inflation risk. \\
        GDP Growth ($G_t$) & Strong economic growth can increase bond yields due to higher demand for capital and expectations of tighter monetary policy. \\
        Unemployment Rate ($U_t$) & Higher unemployment tends to lower bond yields as it signals economic weakness, leading to dovish monetary policy. \\
        Fed Funds Rate ($F_t$) & An increase in the Fed Funds Rate raises short-term yields and influences longer-term bond yields through expectations of future rate hikes. \\
        Money Supply ($M_t$) & A higher money supply can reduce bond yields by increasing liquidity and lowering interest rates. \\
        Consumer Confidence Index ($C_t$) & Higher consumer confidence can increase bond yields as investors anticipate stronger economic activity and higher interest rates. \\
        S\&P 500 Index ($S_t$) & Rising equity markets can lead to higher bond yields as investors shift from bonds to riskier assets, reducing bond prices. \\
        Crude Oil Prices ($O_t$) & Higher oil prices can lead to higher bond yields due to inflationary pressures and expectations of tighter monetary policy. \\
        Gold Prices ($Gd_t$) & Rising gold prices may indicate economic uncertainty, potentially leading to lower bond yields as investors seek safe-haven assets. \\
        US Dollar Index ($D_t$) & A stronger US dollar can put downward pressure on inflation and bond yields by reducing import costs. \\
        INR/USD Exchange Rate ($E_t$) & A depreciating INR (relative to USD) may increase bond yields due to inflationary pressures and capital outflows. \\
        Volatility Index (VIX, $V_t$) & Higher market volatility often leads to lower bond yields as investors move toward safer assets like government bonds. \\
        \bottomrule
    \end{tabular}
    \caption{Effect of Macroeconomic Variables on Bond Yields}
    \label{tab:macro_effects}
\end{table*}

\section*{Methodology} 

This section outlines the methodology employed in this study, which integrates Generative Adversarial Networks (GANs), Reinforcement Learning (RL), and Large Language Models (LLMs) to generate synthetic bond yield data and provide actionable trading insights. The workflow consists of three main components: data preprocessing, synthetic data generation using GANs and RL, and predictive modeling with LLMs.

\subsection*{Data Description}
The dataset used in this study comprises 10 years of monthly bond yield data (2013–2023) for four bond categories: AAA-rated corporate bonds, BAA-rated corporate bonds, US 10-year Treasury notes (US10Y), and high-yield junk bonds. Each bond's yield is influenced by 12 macroeconomic variables that reflect key economic and financial market dynamics. These variables influence bond yields through various economic mechanisms, including interest rate expectations, inflationary pressures, and investor sentiment.

The dataset is structured as follows:
\begin{equation}
\begin{split}
X = \{(&t, Y_{AAA}, Y_{BAA}, Y_{US10Y}, Y_{Junk}, I_t, G_t, U_t,\\
    &F_t, M_t, C_t, S_t, O_t, Gd_t, D_t, E_t, V_t)\}_{t=1}^{120}
\end{split}
\end{equation}
where $t$ represents the month index over 10 years (120 months). The target variables are the bond yields $Y_{AAA}$, $Y_{BAA}$, $Y_{US10Y}$, and $Y_{Junk}$, which respond dynamically to macroeconomic conditions.  

\subsection{Macroeconomic Variables and Their Effect on Bond Yields}

Table \ref{tab:macro_effects} summarizes the expected impact of each macroeconomic variable on bond yields.

\begin{figure*}[!htbp]
    \centering
    \includegraphics[width=\linewidth]{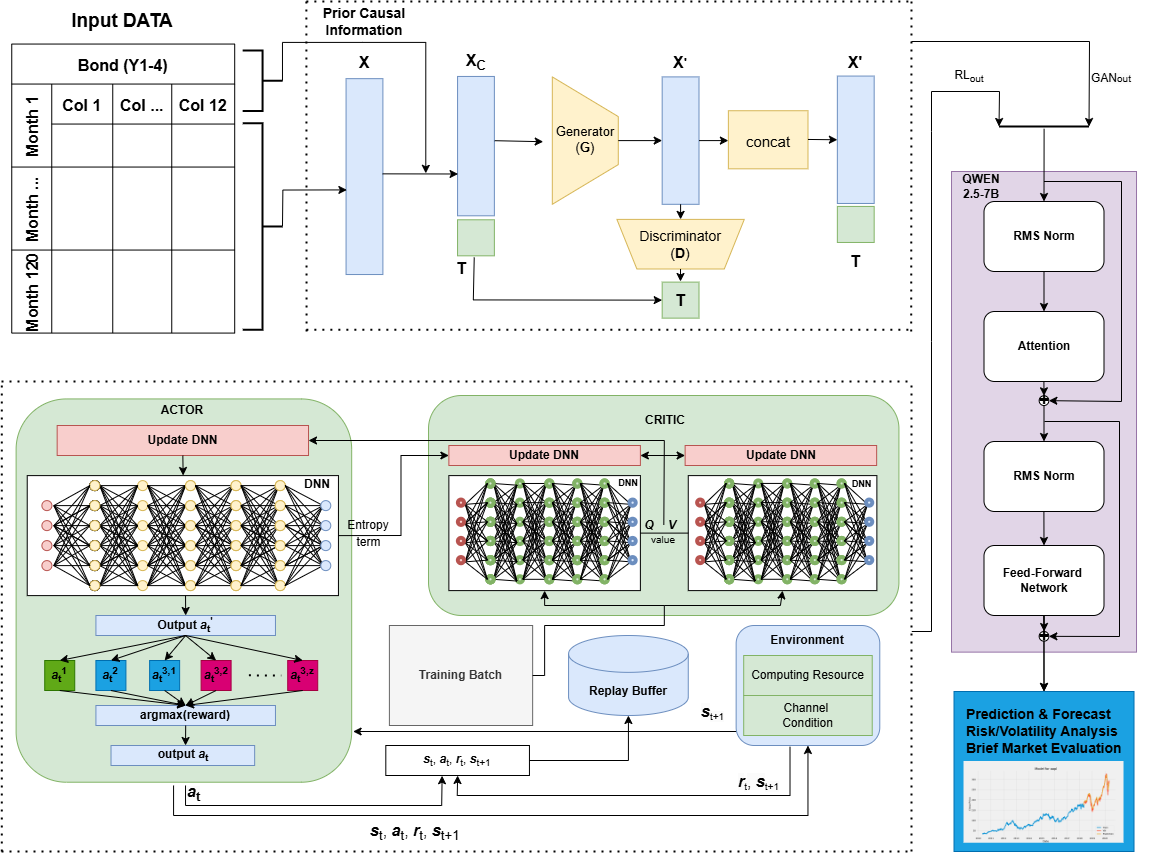}
    \caption{Overall architecture of the models in our pipeline for Causal GANs and Deep Reinforcement Learning with Soft actor Critic as its algorithm.}
    \label{fig:architecture}
\end{figure*}

\subsection*{Synthetic Data Generation}

To address data scarcity and augment the training dataset for predictive modeling, we employ two complementary approaches: GANs and RL. These methods encapsulate the dependencies of the 12 macroeconomic variables on the four bond yields.
\subsubsection*{Causal GAN for Time-Series Data}
We utilize a Causal GAN architecture \cite{yoon2019timegan} to generate synthetic time-series data that preserve temporal dependencies and causal relationships between variables. The Causal GAN consists of a generator $G$, a discriminator $D$, and an embedding network $E$. The generator models the conditional distribution of bond yields given the macroeconomic variables:
\begin{equation}
\begin{split}
G(z \mid X) = \, &\mathbb{P}\Bigl(Y \mid I_t, G_t, U_t, F_t, M_t, C_t, S_t,\\
                 &\quad O_t, Gd_t, D_t, E_t, V_t\Bigr)
\end{split}
\end{equation}
where $z \sim \mathcal{N}(0, 1)$ is a latent noise vector. The discriminator is tasked with distinguishing between real and synthetic data:
\begin{equation}
D(X) = \mathbb{P}(\text{Real} \mid X)
\end{equation}
The training objective minimizes the Wasserstein distance with gradient penalty to stabilize learning:
\begin{equation}
\begin{split}
\min_G \max_D \; &\mathbb{E}_{X_{\text{real}}}\Bigl[D(X_{\text{real}})\Bigr] - \mathbb{E}_{X_{\text{fake}}}\Bigl[D(X_{\text{fake}})\Bigr] \\
&+ \lambda\, \mathbb{E}_{\hat{X}} \Bigl[\Bigl(\|\nabla_{\hat{X}} D(\hat{X})\|_2 - 1\Bigr)^2\Bigr]
\end{split}
\end{equation}
Here, $\hat{X}$ denotes a linear interpolation between real and synthetic samples, and $\lambda$ is a hyperparameter controlling the gradient penalty. This formulation ensures that the generated data maintain realistic statistical properties, including volatility clustering and autocorrelation.

\subsubsection*{Reinforcement Learning with Soft Actor-Critic (SAC)}
\begin{figure}
    \centering
    \includegraphics[width=\linewidth]{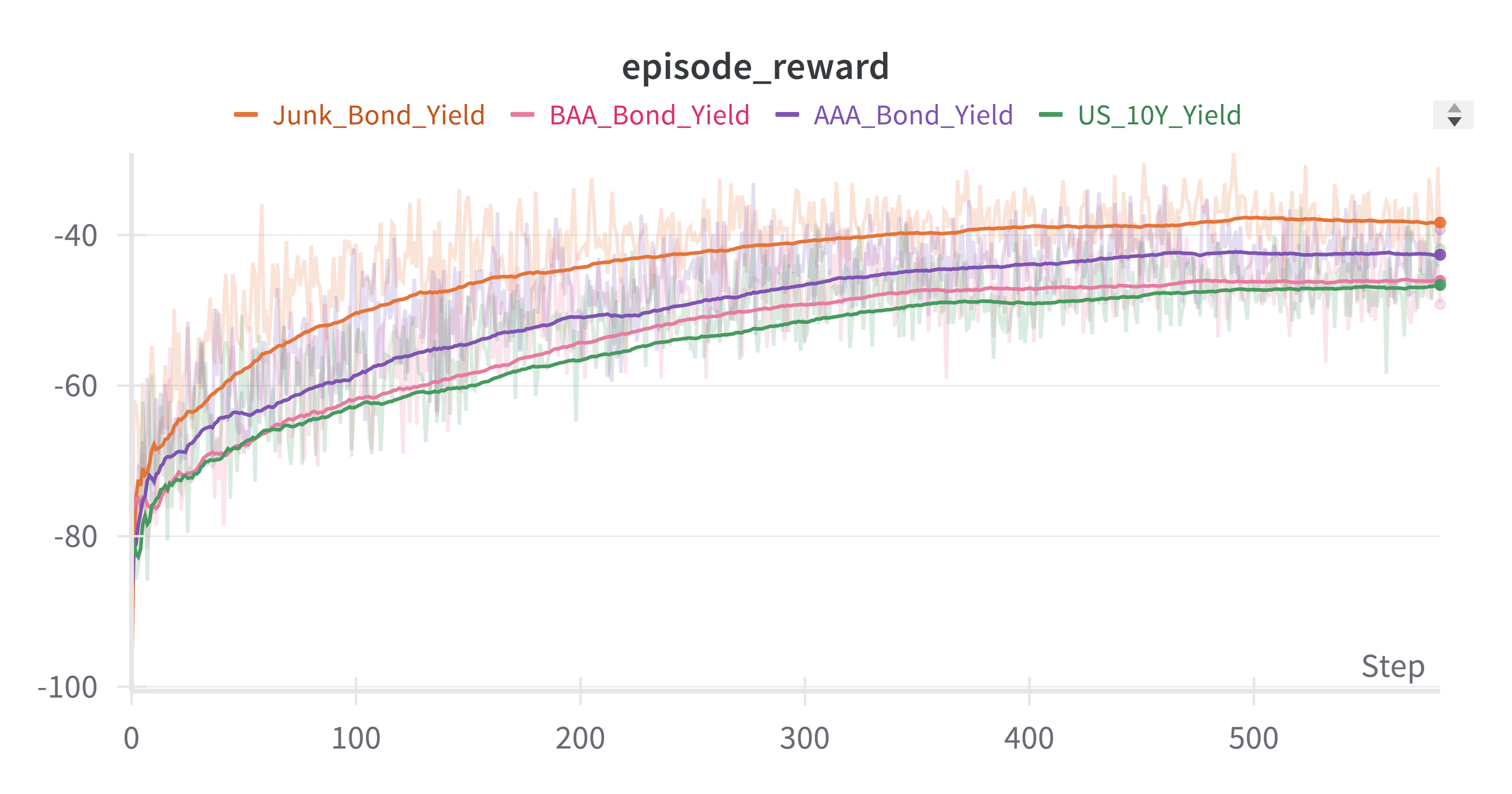}
    \caption{Real-time reward curves for Reinforcement Learning Model}
    \label{fig:rewards}
\end{figure}

To further refine synthetic data generation and capture complex interdependencies, we employ Soft Actor-Critic (SAC) \cite{haarnoja2018soft}, an off-policy RL algorithm grounded in maximum entropy reinforcement learning. SAC optimizes a stochastic policy $\pi_\theta(a \mid s)$ to maximize both expected reward and the entropy of the policy:
\begin{equation}
J(\pi) = \sum_{t=1}^{T} \mathbb{E}_{(s,a) \sim \pi} \Bigl[r(s,a) + \alpha\, \mathcal{H}\bigl(\pi(\cdot \mid s)\bigr)\Bigr]
\end{equation}
In this expression, $r(s,a)$ is a reward function that quantifies the realism of the synthetic data (e.g., via the mean squared error between real and generated bond yields), and $\alpha$ is a temperature parameter controlling the trade-off between exploration and exploitation. The SAC framework comprises three components:
\begin{itemize}
    \item \textbf{Actor Network}: Generates actions $a$, which correspond to adjustments in the synthetic data.
    \item \textbf{Critic Network}: Evaluates state-action pairs $(s,a)$ through Q-values.
    \item \textbf{Entropy Regularization}: Promotes exploration by maximizing the entropy $\mathcal{H}$.
\end{itemize}
The overall SAC loss is formulated as:
\begin{equation}
L_{\text{SAC}} = J_Q + J_\pi + J_\alpha
\end{equation}
where $J_Q$, $J_\pi$, and $J_\alpha$ represent the losses for the critic network, actor network, and the entropy coefficient, respectively. This approach allows the RL agent to interact iteratively with the GAN generator to improve the quality of the synthetic data. Figure \ref{fig:rewards} indicates the reward curves plotted over time for all four bond yield types.

\subsection*{Predictive Modeling with LLMs}
Once synthetic datasets are generated by the GAN and RL models, they are merged with real data and used to train a Large Language Model (LLM) for predictive analytics. Specifically, we employ \textbf{Qwen2.5-7B} \cite{yang2024qwen2} as our predictive model, fine-tuned on both real and synthetic bond data. The LLM takes as input the monthly bond yields $\bigl(Y_{AAA}, Y_{BAA}, Y_{US10Y}, Y_{Junk}\bigr)$ along with the temporal index (month) and outputs three key predictions:

\begin{enumerate}
    \item \textbf{Trading Signals}: BUY/HOLD/SELL recommendations with confidence intervals.
    \item \textbf{Risk Analysis}: Modified Value-at-Risk (VaR) estimates.
    \item \textbf{Volatility Projections}: Conditional variance estimates inspired by GARCH-type models.
\end{enumerate}

The input sequence is defined as:
\begin{equation}
S = \bigl(\text{Month}_t, Y_{AAA,t}, Y_{BAA,t}, Y_{US10Y,t}, Y_{Junk,t}\bigr)_{t=1}^{120}
\end{equation}

The LLM has been fine-tuned using a masked language modeling (MLM) objective:
\begin{equation}
L_{\text{MLM}} = -\sum_{i=1}^{N} \log P(s_i | s_{<i}; \theta)
\end{equation}

By integrating Qwen2.5-7B into our predictive modeling pipeline, we enhance decision-making capabilities, leveraging its contextual understanding to generate interpretable and robust financial forecasts.

\subsection*{Evaluation Framework}
To assess the performance and reliability of our integrated methodology, we employ four evaluation metrics:
\begin{enumerate}
    \item \textbf{LLM-as-Judge}: We employ a seperate LLM, DeepSeek-R1-Distil\cite{deepseek_r1} which evaluates the coherence and decision logic of the trading recommendations on a score of 1-5.
    \item \textbf{Profit/Loss Analysis}: Measures prediction based on accuracy on simulated trading outcomes (1-profit; 0-loss).
    \item \textbf{Mean Absolute Error (MAE)}: Quantifies the deviation between predicted and actual bond yields.
    \item \textbf{Expert Evaluation}: Financial experts qualitatively assess the usefulness of these outputs for practical viability.
\end{enumerate}
This comprehensive evaluation framework ensures that our synthetic data generation and predictive modeling techniques are both quantitatively rigorous and qualitatively sound.

\section*{Experiments and Results}

This section details the experimental setup, presents the results from our synthetic data generation and predictive modeling, and analyzes the performance of our integrated framework using a 4-step evaluation pipeline with automated, human and LLM evaluation. We also compare the utilized frameworks with an ablation study to assess the contribution of individual components.

\begin{table}[h!]
    \centering
    \begin{tabular}{ll}
    \toprule
    \textbf{Model} & \textbf{Size} \\
    \midrule
    CausalGAN & 134K \\
    RL & 6.3M \\
    LLMEval\cite{QwenQwen89:online} & 7B \\
    LLMJudge\cite{deepseek73:online} & 32B \\
    \bottomrule
    \end{tabular}
    \caption{Model Sizes}
    \label{tab:model_sizes}
\end{table}

\subsection*{Experimental Setup}
Our experiments were conducted on a workstation equipped with one NVIDIA \textbf{A100 (40GB)} GPU, running Ubuntu 20.04. The deep learning frameworks used for implementation include PyTorch\cite{paszke2019pytorchimperativestylehighperformance} for GAN and RL training and for running the LLM inferences we use HuggingFace-Transformers
\cite{wolf2020huggingfacestransformersstateoftheartnatural} and DeepEval~\footnote{https://docs.confident-ai.com/}.

\paragraph{Training Hyperparameters:}
\begin{itemize}
    \item \textbf{Causal GAN:} The generator and discriminator networks were trained using the Adam optimizer with a learning rate of \(\eta_{GAN} = 2e-4\) and a batch size of \(8\).
    \item \textbf{Soft Actor-Critic (SAC):} The RL agent used a learning rate of $\eta_{RL} = 1e-4$, with an entropy temperature parameter $\alpha =0.2 $ and a discount factor $\gamma = 0.99$. The actor and critic networks were updated using a replay buffer of size $10000$ and the neural network architecture was defined as $[1024, 1024, 1024]$ with a batch size of $512$ and $\tau = 0.005$.
\end{itemize}

\subsection*{Predictive Performance}
We evaluate the predictive performance of our LLM on both real and synthetic datasets. The LLM has a standardized json output which contains the trading signals (BUY/HOLD/SELL), risk analysis, and volatility projections. Performance is measured using standard metrics including accuracy, mean absolute error (MAE), and profit/loss ratios. 

The MAE metric is computed as:
\begin{equation}
\text{MAE} = \frac{1}{N} \sum_{i=1}^{N} \left| Y_i - \hat{Y}_i \right|
\end{equation}
where \(Y_i\) is the actual bond yield and \(\hat{Y}_i\) is the predicted yield.

\subsection*{Ablation Studies}
Ablation studies were conducted to evaluate the contribution of individual components of our framework. Specifically, we compared:

The performance differences between using solely GAN-generated data, solely RL-refined synthetic data, and the actual data.

\begin{figure*}[!h]
    \centering
    \includegraphics[width=\linewidth]{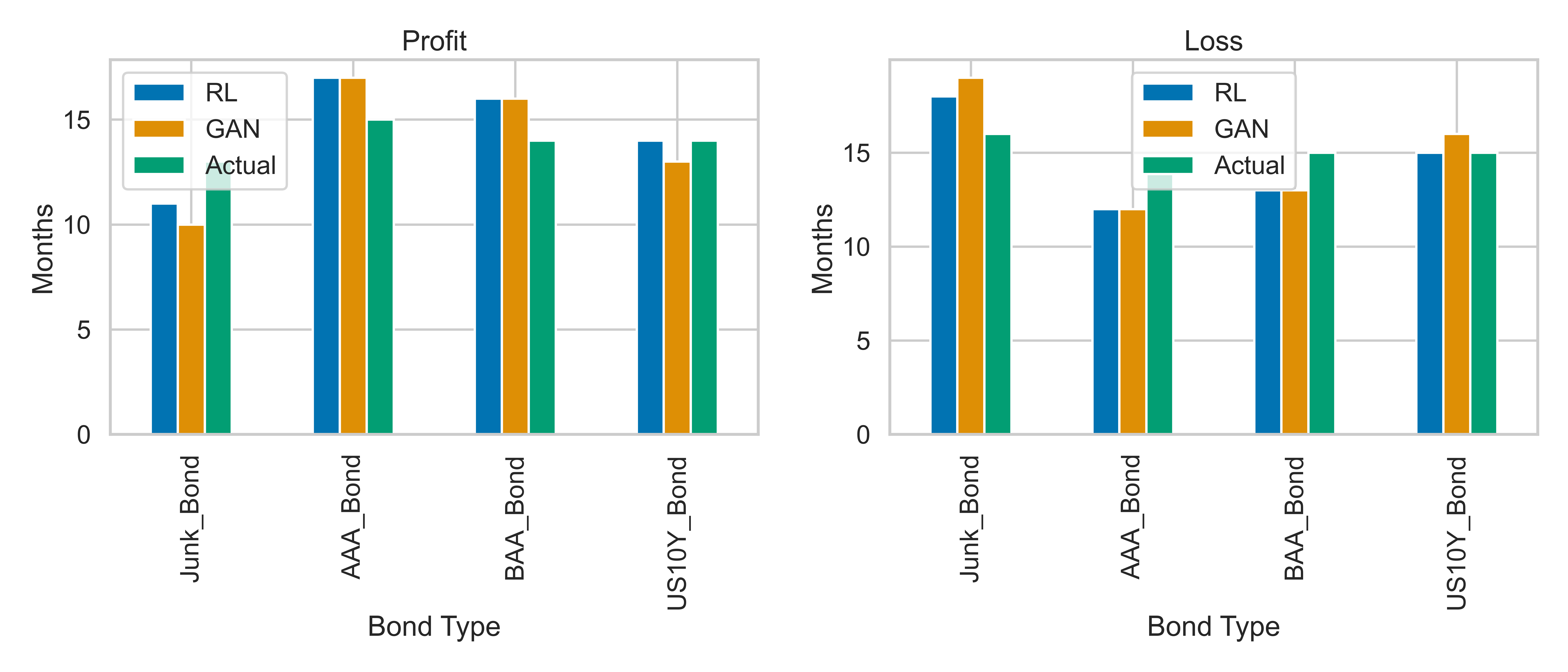}
    \caption{Plots for the Total Profit and Total Loss months between RL, GAN and Actual for each bond type.}
    \label{fig:profitloss}
\end{figure*}

Table \ref{tab:avg_eval_scores} shows the ablation studies on MAE against actual forecasted bond yields. The ablation results indicate that removing the RL component increases the MAE a value of \textbf{0.3} on US 10 year bond yields and \textbf{0.1} on Junk Bond Yields. Figure \ref{fig:profitloss} shows a comparative analysis on the total number of months in which RL/GAN achieves a profit or a loss. 
Our models tend to overestimate actual yields due to the following factors:

\begin{enumerate}
    \item Large Language Models (LLMs) have inherent limitations in accurately predicting future outcomes, particularly in dynamic and uncertain environments.
    \item While our forecasting approach projects higher potential profits, it does not fully account for the significant risk factors associated with market fluctuations and external uncertainties.
\end{enumerate}

\subsection*{Evaluation Results}
We evaluated our integrated methodology using four metrics:

\paragraph{LLM-as-Judge Evaluation:} 

\begin{figure}
    \centering
    \includegraphics[width=\linewidth]{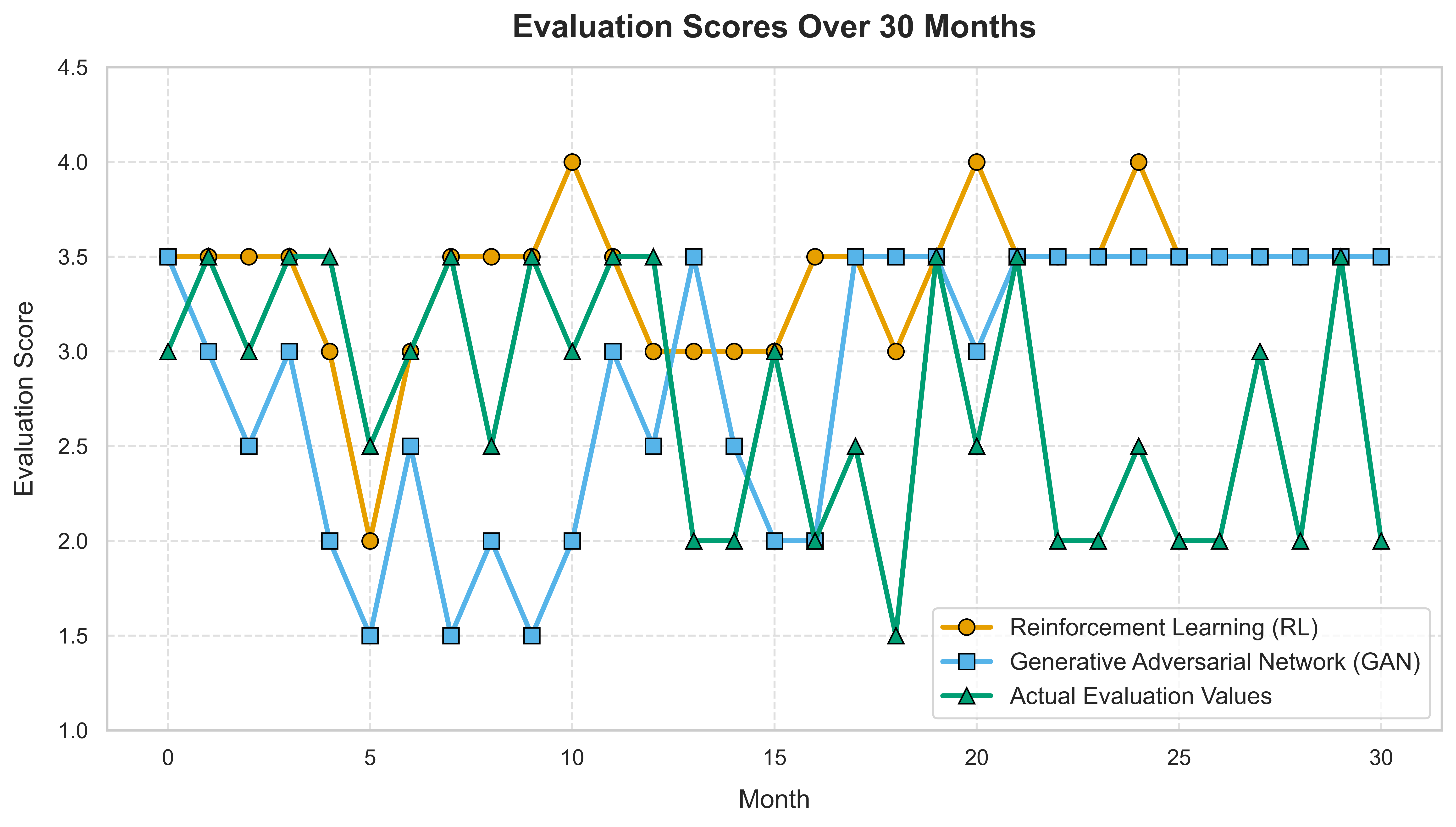}
    \caption{Plot of the evaluation given by the LLM over the last 30 months for each method}
    \label{fig:llmasjudge}
\end{figure}

Figure \ref{fig:llmasjudge} shows the scores given by the LLM as a Judge over a period of 30 months based on a lookback of 2 years. A separate LLM (DeepSeek-R1-Distil) was employed to assess the decision-making quality of the prediction LLM (QWEN2.5-7B) over the last 30 months of data. The judge LLM compares the predicted trading signals with the actual market outcomes in an iterative manner (starting from month 85 to month 120) and assigns scores on a scale of 1 to 5 on a static evaluation criteria. Table \ref{tab:avg_eval_scores} shows that the RL method had the highest average evaluation score of \textbf{3.37}.

\begin{table}[h!]
\centering
\begin{tabular}{lccc}
\toprule
\textbf{Method} & \textbf{Avg. LLM Judge Score}$\uparrow$ \\
\midrule
RL & \textbf{3.37} \\
GAN & 2.87 \\
Actual & 2.58 \\
\bottomrule
\end{tabular}
\caption{Average Evaluation Scores for Different Methods. Best score denoted by \textbf{bold}.}
\label{tab:avg_eval_scores}
\end{table}

\paragraph{Profit/Loss Evaluation:}

A custom evaluation script determined the profitability of the predictions, assigning a value of 1 for profitable predictions and 0 for losses. The overall profit/loss accuracy was computed as:
\begin{equation}
\text{Profit/Loss Accuracy} = \frac{1}{N} \sum_{i=1}^{N} \mathbb{I}\{ \text{Profit}_i \}
\end{equation}

On comparing the number of profit months between RL/Gan and the actual bond yields, the RL- based approach achieves the most number of profit months which was 60\% for AAA and 54\% for BAA Bonds. 

\paragraph{Mean Absolute Error (MAE):}

\begin{table}[h!]
\centering
\begin{tabular}{lll}
\toprule
\textbf{Method} & \textbf{Bond Type} & \textbf{MAE} $\downarrow$ \\
\midrule
GAN & US\_10Y\_Yield & 0.437 \\
GAN & AAA\_Bond\_Yield & 0.343 \\
GAN & BAA\_Bond\_Yield & 0.372 \\
GAN & Junk\_Bond\_Yield & 0.594 \\
RL & US\_10Y\_Yield & \textbf{0.103} \\
RL & AAA\_Bond\_Yield & 0.124 \\
RL & BAA\_Bond\_Yield & 0.174 \\
RL & Junk\_Bond\_Yield & 0.458 \\
\bottomrule
\end{tabular}
\caption{Mean Absolute Error for Different Bond Yields. Best score denoted by \textbf{bold}.}
\label{tab:bond_yields_avg}
\end{table}

The MAE of the predicted bond yields, weighted by their economic impact, was computed using Equation (3). Our analysis shows an average MAE of \textbf{0.2} on all bond yields for our proposed framework. Table \ref{tab:bond_yields_avg} shows the MAE for both RL and GAN against the actual forecasted values. 

\paragraph{Expert Evaluation:}

Three financial experts conducted a qualitative assessment of the model outputs, focusing on the realism of synthetic data and the validity of the trading signals. Their insights and recommendations were summarized into an overall expert evaluation score of \textbf{4.67} for Rl and \textbf{3.67} for GANs. Table \ref{tab:expert_eval_scores} summarises the scores of the expert evaluators. 
\begin{table}[h!]
\centering
\begin{tabular}{lcccc}
\toprule
\textbf{Method} & \textbf{Expert 1} & \textbf{Expert 2} & \textbf{Expert 3} & \textbf{Average}$\uparrow$ \\
\midrule
RL & 4.5 & 5.0 & 4.5 & \textbf{4.67} \\
GAN & 4.0 & 4.0 & 4.5 & 4.17 \\
Actual & 3.5 & 3.5 & 4.0 & 3.67 \\
\bottomrule
\end{tabular}
\caption{Expert Evaluation Scores for Different Methods. Best score denoted by \textbf{bold}.}
\label{tab:expert_eval_scores}
\end{table}

Collectively, these evaluations demonstrate that our integrated framework produces synthetic data that not only replicates the critical market dynamics but also supports predictive performance of forecasting methods.

\section*{Conclusion}

In this paper, we presented an integrated framework for synthesizing financial bond yield data by leveraging a hybrid approach that combines causal Generative Adversarial Networks (GANs) and Soft Actor-Critic (SAC) reinforcement learning, followed by predictive analysis using a fine-tuned Large Language Model: Qwen2.5-7B. Our methodology was designed to encapsulate the intricate dependencies among 12 macroeconomic variables and four bond yields while preserving market-essential statistical properties.

Our extensive experimental results indicate that the synthetic data generated by the proposed framework encapsulates market dynamics of the 12 macroeconomic variables but also closely resembles real-world data, as evidenced by visual comparisons. The predictive performance of the LLM, when trained on the augmented dataset, demonstrated an LLM-as-Judge score of \textbf{3.4}, and a mean absolute error (MAE) of \textbf{0.10} for RL. Additionally, ablation studies confirmed that the reinforcement learning component significantly enhances data quality—its removal increased the MAE by \textbf{0.103}. We show that data points generated by both RL and Causal GANs show significantly better performance in recommendations than the forecasted values of the actual bond yields.

These results demonstrate the potential of our integrated approach in generating high-fidelity synthetic financial data and supporting predictive modeling for bond investment trading and risk management. Future work will focus on further refining the generative models, incorporating additional macroeconomic factors, and extending the framework to other financial instruments and market conditions.


\subsubsection*{Ethical Statement}
This work employs synthetic data generation techniques and advanced machine learning methods to enhance financial forecasting and trading decision support. While our approach mitigates issues of data scarcity and privacy concerns by generating data that closely mimics real-world financial indicators, we acknowledge several ethical considerations.

First, the use of synthetic data must be carefully managed to ensure that the process does not inadvertently propagate biases inherent in historical data. Misrepresentations or oversimplifications in synthetic datasets could lead to flawed financial predictions and decision-making, with potentially adverse economic and societal impacts. Second, although synthetic data help protect individual privacy by avoiding the direct use of sensitive real-world information, there remains a risk of indirect re-identification if the synthetic data are not sufficiently randomized or are combined with external datasets.

Furthermore, as with any AI-driven system, transparency in model development, validation, and deployment is essential. In this study, we incorporate human evaluation and expert assessments as key components of our evaluation framework to ensure the reliability and contextual soundness of our predictions. We are committed to documenting our methodologies and results comprehensively, thereby promoting reproducibility and enabling independent verification of our findings.

Overall, our research aims to contribute positively to financial decision-making processes while actively mitigating potential negative societal implications.

\bibliography{aaai25}

\end{document}